\newcommand{\non}{\nonumber}
\newcommand{\R}{{\mathbb R}}   

\def\T{\tiny\mbox{\rm T}}

\newcommand{\vect}[2]{\left[\begin{array}{c} #1 \\  #2 \end{array} \right]}

\newcommand{\vectf}[4]{\left[\begin{array}{c} #1 \\  #2 \\  #3 \\ #4 \end{array}\right]}

\newcommand{\mtwo}[4]{\setlength\arraycolsep{5pt}\left[\begin{array}{cc}#1 & #2 \\ #3 & #4\end{array}\right]}

\newcommand{\mthree}[9]{\setlength\arraycolsep{5pt}
                        \left[\begin{array}{ccc}#1 & #2 & #3 \\
                                                #4 & #5 & #6 \\
                                                #7 & #8 & #9\end{array} \right]}

\def\bfc{\bm c}

\def\bfu{\bm u}
\def\bfv{\bm v}
\def\bfw{\bm w}
\def\bfx{\bm x}
\def\bfy{\bm y}

\def\bfA{\bm A}
\def\bfB{\bm B}
\def\bfC{\bm C}
\def\bfD{\bm D}

\def\bfI{\bm I}

\def\bfO{\bm O}
\def\bfP{\bm P}
\def\bfQ{\bm Q}
\def\bfR{\bm R}
\def\bfS{\bm S}

\def\bfX{\bm X}

\def\bftheta{\bm\theta}

\def\bfchi{\bm\chi}

\def\bfomega{\bm\omega}
\documentclass[11pt, letterpaper]{article}

\usepackage{graphicx}
\usepackage{amsmath}
\usepackage{amssymb}
\usepackage{setspace}
\usepackage{epstopdf}
\usepackage{color}
\usepackage[letterpaper,left=1in,right=1in,top=1in,bottom=1in]{geometry}
\usepackage[linesnumbered,ruled,vlined,longend]{algorithm2e}
\usepackage{multirow}
\usepackage{longtable}
\usepackage{rotating}
\usepackage{threeparttable}
\usepackage{colortbl}
\usepackage{amsthm}
\usepackage[modulo]{lineno}
\usepackage{enumerate}
\usepackage{bm}
\usepackage{algorithmic}

\newtheorem{remark}{Remark}

\title{\Large \bf Distributed simultaneous state and parameter estimation of nonlinear systems }

\author{\centerline{\normalsize Siyu Liu$^{a,b}$, Xunyuan Yin$^{c}$, Jinfeng Liu$^{b,}$\thanks{Corresponding author: J. Liu. Tel: +1-780-492-1317. Fax: +1-780-492-2881. Email: jinfeng@ualberta.ca}, Feng Ding$^{a}$}\vspace{5mm}\\
    \centerline{\small $^{a}$ School of Internet of Things Engineering, Jiangnan University, Wuxi\ 214122, China}\\
    \centerline{\small $^{b}$ Department of Chemical \& Materials Engineering, University of Alberta,}\\
    \centerline{\small Edmonton, AB, Canada, T6G 1H9}\\
    \centerline{\small $^{c}$ School of Chemical and Biomedical Engineering, Nanyang Technological University,}\\
    \centerline{\small 62 Nanyang Drive, Singapore, 637459}}

\allowdisplaybreaks

\begin{document}

\date{}

\maketitle
\setstretch{1.39}

\begin{abstract}
In this paper, we consider distributed simultaneous state and parameter estimation for a class of nonlinear systems, for which the augmented model comprising both the states and the parameters is only partially observable. Specifically, we first illustrate how the sensitivity analysis (SA) can select variables for simultaneous state and parameter estimation. Then, a community structure detection (CSD) based process decomposition method is proposed for dividing the entire system into interconnected subsystems as the basis of distributed estimation. Next, we develop local moving horizon estimators based on the configured subsystem models, and the local estimators communicate with each other to exchange their estimates. Finally, an SA and CSD based distributed moving horizon estimation (DMHE) mechanism is proposed. The effectiveness of the proposed approach is illustrated using a chemical process consisting of four connected reactors.
\end{abstract}


\section{Introduction} \label{sec:introduction}

State estimation is essential for process modeling, monitoring, control, and fault diagnosis.
Large-scale complex chemical processes are becoming predominant in the operation industry because of their economic benefits. It is worth mentioning that most of the existing state estimation methods were developed within the centralized framework, which can be used to handle small-scale processes, but are not favorable for large-scale processes with more complex structure. The distributed state estimation has been applied to some complex problems in process systems engineering \cite{Yin2019_IEEE_TCST,Mah2016_ACSP}. In particular, different distributed moving horizon estimation algorithms have been proposed and stability analysis has been conducted for this type of algorithms \cite{Zhang2013_JPC,Batt2019_IEEETAC}. For example, a distributed moving horizon estimation (DMHE) was applied for a two-time-scale nonlinear system, which was decomposed into a fast subsystem and several slow subsystems \cite{Yin2017_Auto}. An observer-enhanced DMHE algorithm was developed for the nonlinear system subject to time-varying communication delays \cite{Zhang2014_JPC}. Most research efforts were devoted to the development of new estimation algorithms. However, the fundamental problem of subsystem decomposition has received relatively much less attention. Pourkargar et al. studied the effects of various decompositions on the control performance and computational efficiency \cite{Pour2018_CERD}. They showed that the proper subsystem decomposition may lead to accurate estimates, reduced computational cost, or even improvement of the system observability. Motivated by the mentioned advantages, we propose a systematic decomposition approach based on the community structure detection.

The concept of community structure detection, which originated from the network theory, provides a promising method for solving the subsystem decomposition \cite{TangWT2018_CCE,Fortunato2010_PR}. Unlike the traditional decomposition methods that use the blocks and hierarchical structure \cite{DF2019_JFI_WangLJ,DF2021_ACSP_LiMH,DF2020_JFI_JiY}, the community structure detection divides a large-scale network into communities/groups with more connections within each group than among groups. The modularity function has been used widely as the basis of detecting communities in complex networks \cite{Newman2006_PANS,Fortunato2010_PR}. The state and measured variables are treated as nodes and are connected to each other through directed edges. Yin and Liu developed an initial attempt based on the modularity to decompose nonlinear networks into subsystems for distributed state estimation and control \cite{Yin2019_AIChE}, yet it requires the system to be observable. However, it is challenging to test the observability of nonlinear systems because of the calculation of higher-order Lie derivatives. Instead, some approximation techniques were used, such as the linearization \cite{Nahar2019_CCE} and sensitivity analysis \cite{Grubben2018_IJC} of nonlinear systems.

%
In many first-principle models for complex processes, the values of some parameters are unknown a priori. The accuracy of the parameters affects whether the model can capture the process dynamics. Hence, it is more favorable if the states and parameters are both estimated based on the measured data. There have been some results on simultaneous state and parameter estimation for linear systems \cite{Stoj2020_RNC}, nonlinear systems \cite{DF2020_RNC_LiuSY,DF2021_ND_LiuSY}, and other industrial applications, including sludge wastewater treatment plants \cite{Bezz2013_CEP} and flooding forecasting \cite{Zili2019_JH}. In addition, simultaneous state-parameter estimation is of importance for fault diagnosis and control. For example, the joint robust estimation algorithm was proposed for the stochastic linear systems with sensor, component, and parameter faults \cite{Stoj2020_RNC}. However, most of the existing results have a prominent feature, that is, they require the entire system to be observable. In \cite{Liu2021_IECR}, Liu and coworkers first paid attention to the case when the entire system is only partially observable and discussed the simultaneous estimation by selecting appropriate variables based on the sensitivity analysis. While sensitivity analysis has been devoted to select the model parameters \cite{Kra2013_CCE,Stigter2015_Auto}, how it plays a role in distributed simultaneous estimation has never been discussed.


In this work, we address the simultaneous state and parameter estimation problem when the augmented system is not fully observable, and propose a solution using a distributed framework. The sensitivity analysis is applied to check the system observability and then select which parameters should be estimated most to extract the maximum information based on the available measurements. After that, we propose a systematic framework for determining the optimal decomposition based on the community structure detection, where the states, parameters, and measurement outputs are considered as nodes in a network. By assigning nodes to communities based on the maximum modularity, the subsystem models of distributed state and parameter estimation are established. Although there are some achievements in the decomposition of distributed state estimation, to our knowledge, this paper is the first work that considers the subsystem decomposition for distributed simultaneous state and parameter estimation. Finally, we develop a DMHE algorithm to achieve our goal based on the results of the sensitivity analysis and community structure detection. The contributions of this paper are as follows:
\begin{itemize}

\item Simultaneous state and parameter estimation is addressed in a distributed way when the entire system is not fully observable.

\item The construction of the directed graph considers parameter-state edges and parameter-output edges compared with the existing work, which lays a foundation for distributed simultaneous estimation.

\item Also, we construct a new adjacency matrix that is suitable for distributed simultaneous estimation, in which the state variables, parameters, and measured output variables are all considered as nodes.

\item The distributed moving horizon estimation algorithm which takes advantage of the sensitivity analysis and community structure detection results is proposed for distributed simultaneous estimation.

\end{itemize}

\section{Problem statement and preliminaries}
\label{Section 2}


\subsection{System description}
In this paper, we consider a class of general nonlinear systems described by
\begin{align}
\label{lsy9_2.1a}
 \bfx(t+1)&=f(\bfx(t),\bfu(t),\bftheta),\\
\label{lsy9_2.1b}
 \bfy(t)&=h(\bfx(t),\bftheta),
\end{align}
where $\bfx(t)\in\R^{n_x}$, $\bfu(t)\in\R^{n_u}$ and $\bfy(t)\in\R^{n_y}$ denote the state, input and output vectors of the nonlinear systems, respectively, $\bftheta\in\R^{n_p}$ is the unknown parameter vector, and $f(\cdot)$ and $h(\cdot)$ represent the nonlinear state and output equations, respectively.


\subsection{Problem formulation}

The main objective of this work is to estimate the parameters and states of the system in (\ref{lsy9_2.1a}) and (\ref{lsy9_2.1b}) in a distributed way. Compared to centralized estimation, distributed estimation can reduce the computational burden, increase the fault tolerance and improve the maintenance flexibility. This includes a few subobjectives: 1) To determine the states and parameters that can be estimated based on the measurements. When not all the variables can be estimated simultaneously, how to select the most estimable state and parameter subset based on the given measurements. 2) To decompose the centralized problems into several subproblems. For this part, the main challenge is how the parameters can be divided into different subsystems like states by using the community structure detection. 3) To develop the distributed estimation scheme.


\section{Proposed approach}
\label{Section 3}

\begin{figure}
 \centering
 \includegraphics[width=0.8\hsize]{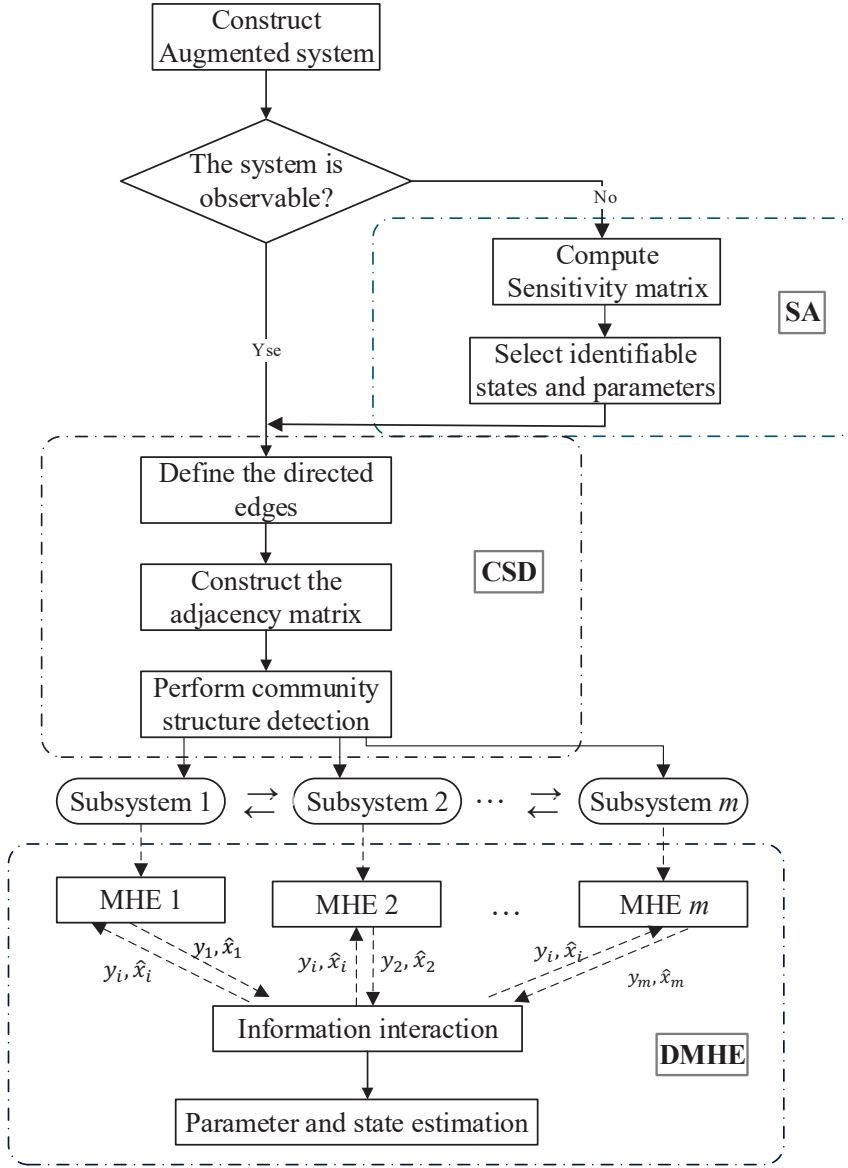}
 \caption{Implementation procedure and information flow of the proposed distributed state and parameter estimation scheme based on SA and CSD}
\label{lsy9_fig1}
\end{figure}

A flow diagram of the proposed procedure is shown in Figure 1. The parameters are first appended to the state vector to construct the augmented system. If the augmented system is observable, the procedure executes the second part: community structure detection (CSD). If not, the sensitivity matrix is constructed and the identifiable states and parameters are selected based on the sensitivity analysis. After determining the estimable states and parameters, the original system is decomposed into several subsystems based on the community structure detection, where the states, parameters and outputs are considered as nodes to define the directed graph and construct the adjacency matrix. Finally, we propose an SA and CSD based distributed moving horizon estimation for distributed simultaneous estimation of the states and parameters.

\subsection{Augmented system construction}

In order to achieve simultaneous estimation, we consider to append the parameters to the state vector. The following augmented system can be obtained:
\begin{align}
\label{lsy9_3.1a}
 \bfx_{\theta}(t+1)&=\vect{f(\bfx(t), \bfu(t), \bftheta(t))}{\bftheta(t)}:=f_{\theta}(\bfx_{\theta}(t), \bfu(t)),\\
\label{lsy9_3.1b}
 \bfy(t)&=h(\bfx(t), \bftheta(t)):=h_{\theta}(\bfx_{\theta}(t)),
\end{align}
where $\bfx_{\theta}(t):=[\bfx^{\T}(t), \bftheta^{\T}(t)]^{\T}\in\R^{n_x+n_p}$ is the augmented state vector, and $f_{\theta}(\cdot)$ and $h_{\theta}(\cdot)$ denote the augmented state equation and output equation, respectively.

When $\bfx_{\theta}(t)$ is observable, many existing state estimation methods can be applied directly. However, in unobservable situations, it will be challenging to obtain accurate estimation results. Therefore, the first step is to test the observability of the augmented state vector $\bfx_{\theta}(t)$ based on given output measurements.


\subsection{Observability of the augmented system}

To check the observability of the system in (\ref{lsy9_3.1a}) and (\ref{lsy9_3.1b}), one common method is to calculate the high-order Lie derivatives of the function $h$ regarding $f$ and their differential. However, it may be computationally expensive when the system order is large. In this case, a practical method is to linearize the nonlinear system at different points along the typical trajectories and check the observability of the linearized models. If the linearized models are observable, it can be concluded that the nonlinear system is locally observable along the considered trajectories.

Consider $N$ sampling points from $t-N+1$ to $t$ along a trajectory of the system in (\ref{lsy9_3.1a}) and (\ref{lsy9_3.1b}). The linearization of the system at each sampling point can be described using the following form with the assumption of zero process and measurement noise (without the loss of generality):
\begin{align}
\label{lsy9_3.2a}
 \bfx_{\theta}(t+1)&=\bfA_{\theta}(t)\bfx_{\theta}(t)+\bfB_{\theta}(t)\bfu(t),\\
\label{lsy9_3.2b}
 \bfy(t)&=\bfC_{\theta}(t)\bfx_{\theta}(t),
\end{align}
where matrices $\bfA_{\theta}(t):=\frac{\partial f_{\theta}}{\partial\bfx_{\theta}}\big|_{\bfx_{\theta}=\bfx_{\theta}(t)}$, $\bfB_{\theta}(t):=\frac{\partial f_{\theta}}{\partial\bfu}\big|_{\bfu=\bfu(t)}$, and $\bfC_{\theta}(t):=\frac{\partial h_{\theta}}{\partial\bfx_{\theta}}\big|_{\bfx_{\theta}=\bfx_{\theta}(t)}$. The observability matrix of Equations (\ref{lsy9_3.2a}) and (\ref{lsy9_3.2b}) at each sampling point is as follows:
\[
 \bfO(t)=\vectf{\bfC_{\theta}(t-N+1)}{\bfC_{\theta}(t-N+2)\bfA_{\theta}(t-N+1)}{\vdots}{\bfC_{\theta}(t)\bfA_{\theta}(t-1)\cdots\bfA_{\theta}(t-N+1)}.
\]
If $\bfO(t)$ is full rank, the nonlinear system is locally observable along the trajectory \cite{Prol2017}.

\subsection{Sensitivity matrix}

To quantitatively investigate parameter effects, we introduce the sensitivity analysis. The sensitivity of the output $\bfy(t)$ with respect to the parameter $\bftheta$ is defined as:
\[
 \bfS_{y,\theta}(t):=\frac{\partial \bfy(t)}{\partial\bftheta}.
\]
Similarly, the sensitivity of $\bfy(t)$ with respect to the initial state $\bfx(0)$ is represented as $\bfS_{y,x(0)}(t):=\frac{\partial \bfy(t)}{\partial\bfx(0)}$. In order to calculate two sensitivities, we first define the sensitivity of the state to the parameter as $\bfS_{x,\theta}(t):=\frac{\partial \bfx(t)}{\partial\bftheta}$. According to the nonlinear model shown in (\ref{lsy9_2.1a})--(\ref{lsy9_2.1b}), $\bfS_{y,\theta}(t)$ can be computed by solving the following two equations
\begin{align}
\label{lsy9_3.3a}
 \bfS_{x,\theta}(t+1)&=\frac{\partial f}{\partial\bfx}(t)\bfS_{x,\theta}(t)+\frac{\partial f}{\partial\bftheta}(t),\\
\label{lsy9_3.3b}
 \bfS_{y,\theta}(t)&=\frac{\partial h}{\partial\bfx}(t)\bfS_{x,\theta}(t)+\frac{\partial h}{\partial\bftheta}(t),
\end{align}
with the initial condition $\bfS_{x,\theta}(0)=\bf0$.

To obtain another sensitivity $\bfS_{y,x(0)}(t)$, the initial state $\bfx(0)$ can be considered as a virtual parameter of the system. Define the sensitivity of the state to the initial condition as $\bfS_{x,x(0)}(t):=\frac{\partial\bfx(t)}{\partial\bfx(0)}$. The sensitivity $\bfS_{y,x(0)}(t)$ can be calculated by solving the following equations:
\begin{align}
\label{lsy9_3.3f}
 \bfS_{x,x(0)}(t+1)&=\frac{\partial f}{\partial\bfx}(t)\bfS_{x,x(0)}(t),\\
\label{lsy9_3.3g}
 \bfS_{y,x(0)}(t)&=\frac{\partial h}{\partial\bfx}(t)\bfS_{x,x(0)}(t),
\end{align}
with the initial value $\bfS_{x,x(0)}(0)=\bfI$. It is obvious that there are no terms $\frac{\partial f}{\partial\bfx(0)}(t)$ and $\frac{\partial h}{\partial\bfx(0)}(t)$ because $\bfx(0)$ is not explicitly represented in $f(\cdot)$ and $h(\cdot)$.

Based on Equations (\ref{lsy9_3.3f}) and (\ref{lsy9_3.3g}), the sensitivity $\bfS_{y,x_{\theta}(0)}(t)$ of the linearized system in (\ref{lsy9_3.2a}) and (\ref{lsy9_3.2b}) can be rewritten in the following at the sampling point $t$:
\begin{align}
 \bfS_{y,x_{\theta}(0)}(t)&=\frac{\partial h_{\theta}}{\partial\bfx}(t)\frac{\partial f_{\theta}}{\partial\bfx}(t-1)\frac{\partial f_{\theta}}{\partial\bfx}(t-2)\cdots\frac{\partial f_{\theta}}{\partial\bfx}(0)\non\\
              &=\bfC_{\theta}(t)\bfA_{\theta}(t-1)\bfA_{\theta}(t-2)\cdots\bfA_{\theta}(0).\non
\end{align}
For each sampling time $t$, we can collect the most recent $N$ sensitivities $\bfS_{y,x_{\theta}(t-N+1)}(i)$, $i=t, t-1, \ldots, t-N+1$, to form a sensitivity matrix $\bfS_{\theta}(t)$ as follows, and test the rank of the following sensitivity matrix along a typical trajectory in a data window of the augmented system:
\begin{align}
 \bfS_{\theta}(t)=\vectf{\bfS_{y,x_{\theta}(t-N+1)}(t-N+1)}{\bfS_{y,x_{\theta}(t-N+1)}(t-N+2)}{\vdots}{\bfS_{y,x_{\theta}(t-N+1)}(t)},
\end{align}
where $N$ is the data window length. The sensitivities of $\bfy(i)$ with respect to the augmented state $\bfx_{\theta}(t-N+1)$ are calculated by
\begin{align}
 \bfS_{y,x_{\theta}(t-N+1)}(i)=&\frac{\partial h_{\theta}}{\partial\bfx_{\theta}}(i)\frac{\partial f_{\theta}}{\partial\bfx_{\theta}}(i-1)\frac{\partial f_{\theta}}{\partial\bfx_{\theta}}(i-2)\cdots\frac{\partial f_{\theta}}{\partial\bfx_{\theta}}(t-N+1),\non
\end{align}
where
\begin{equation}
 \frac{\partial f_{\theta}}{\partial\bfx_{\theta}}(i)=\mtwo{\frac{\partial f}{\partial\bfx}(i)}{\frac{\partial f}{\partial\bftheta}}{\bf0}{\bfI_{n_p\times n_p}}, \quad
 \frac{\partial h_{\theta}}{\partial\bfx_{\theta}}(i)=\left[\frac{\partial h}{\partial\bfx}(i) \ \ \frac{\partial h}{\partial\bftheta}\right].\non
\end{equation}
\begin{remark}
According to the derivations of $\bfO(t)$ and $\bfS_{\theta}(t)$, we can find that $\bfO(t)$ and  $\bfS_{\theta}(t)$ include the same information from $t-N+1$ to $t$. Therefore, The rank of $\bfS_{\theta}(t)$ has been used as an indication of the observability of nonlinear systems. If the sensitivity matrix $\bfS_{\theta}(t)$ is full rank along all of the sampling points and is well-conditioned, the augmented state vector $\bfx_{\theta}(t)$ can be estimated using the input and output data.
\end{remark}

\subsection{Variable selection}
\label{orthogonalization}

When $\bfS_{\theta}(t)$ is not full rank along all the sampling points or is ill-conditioned, this means that not all the variables in the augmented state vector $\bfx_{\theta}(t)$ can be estimated. In view of this situation, how to estimate $\bfx_{\theta}(t)$ is more challenging. Increasing the number of the output variables to make $\bfS_{\theta}(t)$ full rank and well-conditioned may be a solution. In this work, we consider the case of not increasing the output variables, and focus on resorting to the information in $\bfS_{\theta}(t)$ to select those variables that are important for the output prediction to estimate. The larger the sensitivity of one parameter, the more sensitive the system response with respect to small perturbations of this parameter.

There are several typical methods for selecting the parameters based on the sensitivity matrix: The first one is the correlation method \cite{Miao2011_SIAM}, and its central idea is to find unidentifiable parameters by studying the linear correlation of the columns of the sensitivity matrix. However, when the method detects a pair of correlated parameters or even over one pair of correlated parameters, there is no criterion in which one parameter from a pair is unidentifiable. The principal component analysis (PCA) based method \cite{Degen2004_JPC} has been developed to rank the influence of parameters on the model output first based on the criterions and then these parameters are determined as identifiable or unidentifiable. Due to not a single criterion used by the PCA-based method, it does not select a single parameter, but a set of parameters. Thus, the PCA-based method does not find some parameters that make the sensitivity matrix rank deficient. In addition, the orthogonalization method \cite{Yao2003,Chu2012_AIChE} and the eigenvalue method \cite{Quaiser2009} were proposed to select the parameter subset. The idea of orthogonalization method is to check the linear dependencies of columns of the sensitivity matrix. Unlike the correlation method, the orthogonal method does not calculate the correlation between different columns of $\bfS_{\theta}(t)$. Instead, the perpendicular distance of one column to the vector space spanned by the other columns is calculated as a measurement of the linear dependency. Therefore, the sensitivity of the output to the parameters and the dependence of the parameters can be calculated at the same time to select the identifiable parameters. The eigenvalue method used the same principle as the orthogonalization method, but it discards the unidentifiable parameters.

From the comparisons of the four methods, we apply the orthogonalization method to select the variables. The procedure for sequentially selecting the most important parameters for simultaneous state and parameter estimation using the orthogonalization method is as follows:
\begin{enumerate}
\item
At time $t$, calculate the norm of each column of the normalized $\bfS_{\theta}(t)$.
\item
Initialize $l=1$, select the column with the largest norm, and denote it as $\bfX_l$.
\item
Project the sensitivity vectors of the unselected parameters onto the space orthogonal to the space formed by the sensitivity vectors of the previously selected parameters, and compute the orthogonal projection matrix that cannot be expressed by $\bfX_l$: $\bfP_l=\bfI-\bfX_l(\bfX_l^{\T}\bfX_l)^{-1}\bfX_l^{\T}$.
\item
Calculate the norm of each column of the residual matrix: $\bfR_l=\bfP_l\bfS_{\theta}(t)$, select the column from $\bfS_{\theta}(t)$ that corresponds to the column with the largest norm in $\bfR_l$, and add it to $\bfX_l$ as a new column to form $\bfX_{l+1}$.
\item
If the largest norm of the columns in $\bfR_l$ is larger than a prescribed cutoff value, go to Step 3 with $l:=l+1$; otherwise, obtain the selected elements in $\bfx_{\theta}(t)$ corresponding to the selected columns in $\bfS_{\theta}(t)$, terminate this process.

\end{enumerate}

\begin{remark}
The key component of this procedure is the orthogonalization projection used in Step 3. The projection aims to remove the parameter's effect on the output covered by the selected parameters. In each step, we select the parameter which has the largest not yet covered effect.
\end{remark}

\subsection{Definition of directed edges }
\label{edge}

The sensitivity analysis determines which parameters and states can be estimated, such that the following system decomposition can be carried out. From the perspective of community structure detection, the original nonlinear system can be viewed as a large network, with state, parameter, and output variables being vertices in the network. In the community structure detection part of the proposed procedure is shown in Figure~\ref{lsy9_fig1}, a directed graph is used first to determine the connectivity between state, parameter, and output variables. All these variables selected by the sensitivity analysis are considered as the nodes in the directed graph. Let $y_j$, $\theta_p$, and $x_i$ denote the $j$-th, $p$-th, and $i$-th element of the vectors $\bfy$, $\bftheta_s$ and $\bfx_s$, ($\bftheta_s$ and $\bfx_s$ are constructed by the selected parameters and states), where $i=1,\ldots,n_{\bar{x}}$, $p=1,\ldots,n_{\bar{p}}$, and $j=1,\ldots,n_{\bar{y}}$. $f_i$ denotes the $i$-th element of $f$ and $h_j$ denotes the $j$-th element of $h$. The edges are placed by using the following rules:
\begin{itemize}
\item State-state edge: an edge from $x_i$ to $x_k$ if $\partial f_k(\bfx,\bfu,\bftheta)/\partial x_i\neq0$.


\item Parameter-state edge: an edge from $\theta_p$ to $x_k$ if the parameter directly affects the state.

\item State-output edge: an edge from $x_i$ to $y_j$ if $\partial h_j(\bfx,\bftheta)/\partial x_i\neq0$.

\item Parameter-output edge: an edge from $\theta_p$ to $y_j$ if the parameter directly affects the output.
\end{itemize}
\begin{remark}
It is worth noting that the rules used for constructing directed edges among nodes in this work deffer from those adopted in the existing work \cite{Yin2019_AIChE,Zhang2019_CERD}. Specially, not only the state-state edges and state-output edges are considered, but the edges that connect the parameters and states, parameters and outputs are also constructed. This is due to considering simultaneous estimation of parameters and states, rather than only state estimation problems. Therefore, we use directed edges to capture the connections between parameters and states and between parameters and outputs.
\end{remark}

\begin{figure}[!hbt]
 \centering
 \includegraphics[width=0.8\hsize]{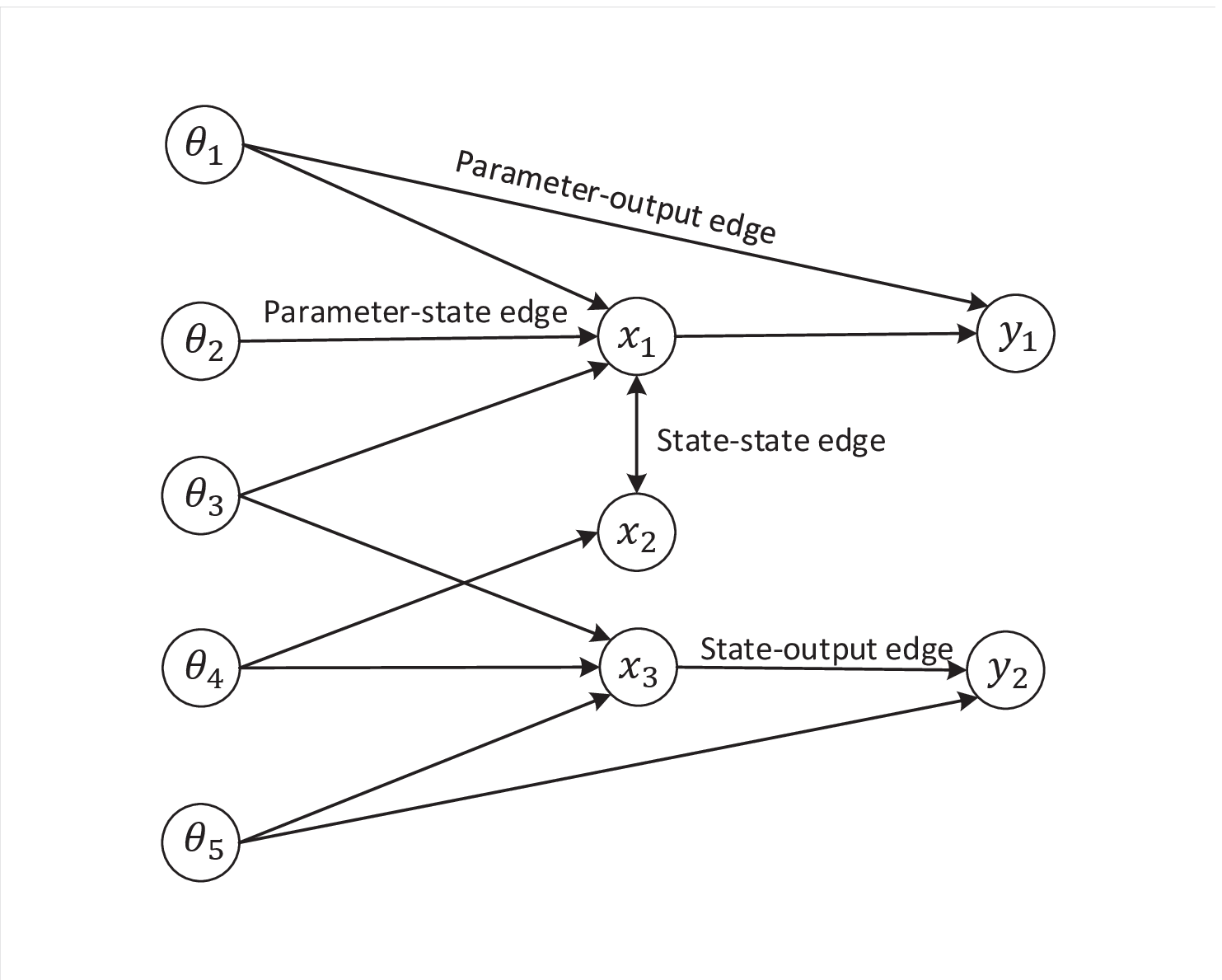}
 \caption{A directed graph example}
\label{lsy9_fig_directed}
\end{figure}

Figure~\ref{lsy9_fig_directed} shows an example directed graph with three states, two outputs, and five parameters. This can be considered as a nontrivial extension of the one defined in \cite{Jogwar2017_CES}. The difference from \cite{Jogwar2017_CES} is that we also consider the parameters as nodes in the directed graphs.

\subsection{Adjacency matrix construction}
\label{Adjacency}

In this work, we propose a systematic approach based on the concept of community structure detection, which will be used for the subsystem decomposition, making the connections within each group denser and the connections between group sparser. Here, we resort to measure the modularity to quantitatively assess the quality of different community structures.

Consider a network of $n$ nodes. The modularity of a community structure can be defined as
\begin{equation}
\label{lsy9_3.6a}
 \Omega=\frac{1}{m}\sum_{i=1}^n\sum_{j=1}^n\left(\bfA_{ij}-\frac{k_i^{\rm in}k_j^{\rm out}}{m}\right)\delta(c_i,c_j),
\end{equation}
where $m$ is the total number of edges in the directed graph, $c_i$ is the community to which node $i$ is assigned, $i=1,\ldots,n$, and $\delta(c_i,c_j)$ yields one if nodes $i$ and $j$ belong to the same community (i.e., $c_i=c_j$), $\delta(c_i,c_j)=0$ otherwise, $\bfA_{ij}$ is the $i$-th element of the $j$-th column of the adjacency matrix, $\bfA_{ij}=1$ if there is a directed edge from node $j$ to node $i$ and $\bfA_{ij}=0$ otherwise, the in-degree $k_i^{\rm in}$ denotes the number of edges entering the node $i$ and can be computed by $k_i^{\rm in}=\sum_{j=1}^n\bfA_{ij}$, the output-degree $k_j^{\rm out}$ represents the number of edges leaving the node $j$ and is calculated by $k_j^{\rm out}=\sum_{i=1}^n\bfA_{ij}$. It is noted that the self-loops (the directed edge from a node to itself) are not taken into account, that is, the diagonal elements of the adjacency matrix are always zero.

The modularity values can range between 0 and 1, in which the larger the value of $\Omega$, the better the community structure. The problem of finding the optimal community structure is equivalent to maximizing the modularity value over all community structure candidates. A modularity value between 0.3 and 0.7 indicates a good community structure \cite{Clauset2004}.

In the following, we construct the adjacency matrix for subsystem decomposition. To achieve both the parameter estimation and state estimation, all the states, parameters and outputs are considered as nodes. There are $n_o$ ($n_o=n_{\bar{x}}+n_{\bar{p}}+n_{\bar{y}}$) nodes in the directed graph constructed based on the method in Section \ref{edge}. Let $\bfc_o$ be an augmented vector: $\bfc_o=[x_1,\ldots,x_{n_{\bar{x}}},\theta_1,\ldots,\theta_{n_{\bar{p}}},y_1,\ldots,y_{n_{\bar{y}}}]^{\T}$. The $i$-th element of $\bfc_o$ corresponds to the $i$-th node in the directed graph. When $n_o$ is small, we can construct the adjacency matrix $\bfA$ by examining the existence of the directed edges among the nodes in $\bfc_o$. However, when $n_o$ is not small, it may be difficult to construct $\bfA$ in this way. Alternatively, $\bfA$ can be constructed by the Jacobian matrices of the vector fields $f$ and $h$ in Equations (\ref{lsy9_2.1a})--(\ref{lsy9_2.1b}). Define the following matrices obtained by taking the derivatives at the equilibrium point of the system as
\begin{align}
 \tilde{\bfA}=\frac{\partial f(\bfx,\bfu,\bftheta)}{\partial \bfx}\bigg|_{(\bfx_s,\bftheta_s)}, \quad
 \tilde{\bfB}=\frac{\partial f(\bfx,\bfu,\bftheta)}{\partial \bftheta}\bigg|_{(\bfx_s,\bftheta_s)},\quad
 \tilde{\bfC}=\frac{\partial h(\bfx,\bftheta)}{\partial \bfx}\bigg|_{(\bfx_s,\bftheta_s)}, \quad
 \tilde{\bfD}=\frac{\partial h(\bfx,\bftheta)}{\partial \bftheta}\bigg|_{(\bfx_s,\bftheta_s)}.
\end{align}
Thus, a matrix is construct as follows:
\begin{equation}
 \bfA_o=\mthree{\tilde{\bfA}}{\tilde{\bfB}}{{\bf0}_{n_x\times n_y}}
                      {{\bf0}_{n_p\times n_x}}{{\bf0}_{n_p\times n_p}}{{\bf0}_{n_p\times n_y}}
                      {\tilde{\bfC}}{\tilde{\bfD}}{{\bf0}_{n_y\times n_y}}_{n_o\times n_o}.\non
\end{equation}
The adjacency matrix $\bfA$ can be obtained according to $\bfA_o$ by substituting all nonzero elements in $\bfA_o$ by one, and then setting all the diagonal elements in $\bfA_o$ to be zero.

\begin{remark}
For the system considered in this paper, most of the parameters are constant. When constructing the adjacency matrix, the method of taking partial derivatives of parameters is used to find out whether a certain parameter has an influence on a certain state. Assuming that the parameters change, if the state changes accordingly, it means that there is an impact, otherwise there is no impact.
\end{remark}

\begin{remark}
In this paper, the adjacency matrix constructed for subsystem decomposition is an extension of the method in \cite{Siljak1991}. The difference is that the adjacency matrix in our work is constructed for a nonlinear state-space system. Compared with the adjacency matrix in \cite{Yin2019_AIChE}, the different is that we also consider the parameters as nodes for constructing the adjacency matrix. Combining the states, parameters and measurement outputs into an adjacency matrix enables us to handle distributed simultaneous state estimation and parameter identification at the same time.
\end{remark}

\subsection{Subsystem decomposition based on the community structure detection}

%
Based on the constructed adjacent matrix, we can perform the community structure detection (CSD) for the subsystem decomposition through finding a higher modularity value. For the purpose of reducing the computational burden of the modularity maximization, the fast unfolding algorithm first proposed in \cite{Blondel2008} is adopted in our paper due to its ease of implementation. In the initial community structure, we assign each node to a different community. Thus, the number of communities is equal to the number of nodes. This method is performed iteratively in following two steps: (a) modularity maximization, and (b) community aggregation. First, each node is selected and moved to each neighboring community in order to calculate whether moving the node will increase the modularity value. The node is placed in the community such that the maximum modularity is obtained. Second, it aggregates all nodes belonging to the same community into one node, where the number of nodes is the number of communities now. The implementing of this method is shown in Algorithm~\ref{alg_fastunfolding}:
%

\begin{algorithm} \label{alg_fastunfolding}
    \caption{The steps to detect the communities for subsystem decomposition}

    \textbf{Initialize}: $n_0$ nodes, $n_0$ communities. \\
    \textbf{Maximization}:\\
    Put each node in its own community.\\
    \For{all nodes}{
        Move each node in its adjacent communities including its own which maximizes the modularity value.}
    Update the modularity $\Omega$ and obtain a local maximum.\\
    \textbf{Aggregation}:\\
    All nodes in a community are aggregated into one node. \\
    Go to Maximization step.\\
    \textbf{until} There are no changes and a maximum $\Omega$ is obtained.
\end{algorithm}
It is assumed after subsystem decomposition, a subsystem can be described as follows: 
\begin{align}
\label{lsy9_2a}
    \bfx_{\theta}^i(t+1)&=f_{\theta}^i(\bfx_{\theta}^i(t),\bfu^i(t),\bfw^i(t))+\tilde{f}^i(\bfchi_{\theta}^i(t)),\\
\label{lsy9_2b}
    \bfy^i(t)&=h_{\theta}^i(\bfx^i(t))+\bfv^i(t),
\end{align}
where $\bfx_{\theta}^i(t)\in\R^{n_{x^i}}$, $\bfu^i(t)\in\R^{n_{u^i}}$ and $\bfy^i(t)\in\R^{n_{y^i}}$, $i\in\{1,2,\cdots,m\}$, are the augmented state vector, input and output vectors of the $i$-th subsystem, $\bfw^i(t)\in\R^{n_{w^i}}$ and $\bfv^i(t)\in\R^{n_{v^i}}$ characterize disturbances and measurement noise of subsystem $i$, respectively, $\bfchi^i(t)\in\R^{\chi^i}$ symbolizes the interaction of subsystem $i$ with other subsystems and contains the states in the neighboring subsystems that affects $\bfx^i(t)$. Specifically, $\bfchi^2(t)=[\bfx^{1\T}(t),\bfx^{3\T}(t)]^{\T}$ means the dynamics of the subsystem 2 is also affected directly by the states of the subsystems 1 and 3.
\begin{remark}
In this work, the number of the communities is not fixed, and the objective is to detect the subsystem structures of the maximum $\Omega$. This differs slightly from the existing community-based methods \cite{Zhang2019_CERD} where the number of the communities (subsystems) is predetermined. Obviously, the way of presetting the number of communities may affect the optimality of the decomposition to a certain extent from the perspective of the modularity value.
\end{remark}

\begin{remark}
Following the algorithm in the previous subsections, we obtain the subsystem configurations with the maximum modularity value. Before performing distributed simultaneous estimation, we choose to perform subsystem observability test for the subsystem configuration candidates. If the subsystem observability test is not passed by the configured subsystems, the corresponding subsystem configuration is discarded. Then choose other subsystem configurations with the next largest modularity value and perform this test again until a suitable configuration is found.
\end{remark}

\subsection{DMHE integrated with SA and CSD results }

In this subsection, the distributed simultaneous state and parameter estimation will be developed by using the distributed moving horizon estimation (DMHE) based on the decomposed subsystem models. A schematic diagram that describes the DMHE scheme is given in the third part of Figure~\ref{lsy9_fig1}. Specifically, at time $t$, after selecting variables according to the sensitivity matrix $\bfS_{\theta}(t)$, it determines that the elements of $\bfx_{\theta}(t)$ can be estimated based on the input-output data from $t-N$ to $t$. In the proposed MHE design for each subsystem, the estimation window used in each MHE is considered being the same as the data window length $N$ used in $\bfS_{\theta}(t)$. The design of the proposed distributed MHE for each subsystem $i$, $i=1,2,\ldots$, at time $t$ based on the augmented system (\ref{lsy9_3.1a})--(\ref{lsy9_3.1b}) is described as follows:
\begin{align}
\label{lay9_DMHE1}
 \min \limits_{\tilde{x}^i_{\theta}(t-N),\ldots,\tilde{x}^i_{\theta}(t)}\bigg\{\sum^{t-1}_{l=t-N}\|\bfw_{\theta}^i(l)\|^2_{\bfQ_i^{-1}}+\sum^{t}_{j=t-N}\|\bfv^i(j)\|^2_{\bfR_i^{-1}}+\Gamma^i(\tilde{\bfx}^i_{\theta}(t-N))\bigg\},
\end{align}
\begin{align}
\label{lay9_DMHE2}
 {\rm s.t.} & \tilde{\bfx}^i_{\theta}(l+1)=f^i_{\theta}(\tilde{\bfx}^i_{\theta}(l),\bfu^i(l),\bfw^i_{\theta}(l))+\tilde{f}^i(\hat{\bfchi}_{\theta}^i(t)),\\
\label{lay9_DMHE3}
 &y^i(l)=h^i_{\theta}(\hat{\bfx}^i_{\theta}(l))+\bfv^i(l),\\
\label{lay9_DMHE4}
 &\tilde{\bfx}^i_{\theta}(l)\in\mathbb{X}^i_{\theta},\quad \bfv^i(l)\in\mathbb{V}^i,\quad l=t-N,\ldots,t,\\
\label{lay9_DMHE5}
 &\bfw_{\theta}^i(l)\in\mathbb{W}^i_{\theta},\quad l=t-N,\ldots,t-1,\\
\label{lay9_DMHE6}
 &\tilde{\bfx}^i_{\theta,u}(t-N)=\tilde{\bfx}^i_{\theta,u}(t-N|t-1),\  u\in U(t),\\
\label{lay9_DMHE7}
 &\bfw^i_{\theta,u}(l)=0,\  u\in U(t), \ l=t-N,\ldots,t-1,
\end{align}
where $\tilde{\bfx}^i_{\theta}$ is the prediction of $\bfx^i_{\theta}$ within the optimization problem, $N$ is the
estimation horizon, $\hat{\bfx}^i_{\theta}$, $\bfQ_i$ and $\bfR_i$ denote the covariance matrices of $\bfw^i$ and $\bfv^i$, respectively, $\hat{\bfchi}_{\theta}$ represents the estimate obtained by other MHEs. Note that the subsystems communicate with each other and the communicated information is used in interactive compensation.

Once the optimization problem (\ref{lay9_DMHE1})--(\ref{lay9_DMHE7}) is solved, a series of solution is determined as $\{\tilde{x}^{i*}_{\theta}(t-N),\ldots,\tilde{x}^{i*}_{\theta}(t)\}$, of which the last element $\tilde{x}^{i*}_{\theta}(t)$ is adopted as the current optimal estimate and denoted as $\hat{x}^i_{\theta}(t)$.

In the optimization problem (\ref{lay9_DMHE1})--(\ref{lay9_DMHE7}), Equation (\ref{lay9_DMHE1}) is the cost function for each MHE estimator, and the last term $\Gamma^i(\tilde{\bfx}^i_{\theta}(t-N))$, called the arrival cost in the MHE, summarizes the previous information of the measurements before the current window. However, an algebraic expression for the arrival cost only exists sometimes, such as the linear unconstrained case. Hence, we choose a quadratic arrival cost with a constant weighting matrix as an approximated arrival cost:
\[
 \Gamma(\tilde{\bfx}^i_{\theta}(t-N))=\|\tilde{\bfx}^i_{\theta}(t-N)-\tilde{\bfx}^i_{\theta}(t-N|t-1)\|^2_{\bfP_i^{-1}},
\]
where $\bfP$ is a constant weighting matrix and $\tilde{\bfx}^i_{\theta}(t-N|t-1)$ is the estimate of $\tilde{\bfx}^i_{\theta}(t-N)$ obtained at the previous time instant $t-1$. It has been shown that, under quite general assumptions, such a simple arrival cost suffices to ensure the convergence of the algorithm, provided that $\bfP$ is adequately chosen to avoid an overconfidence on the existing estimates \cite{Farina2010_Auto,A2008_Auto}. Equations (\ref{lay9_DMHE2}) and (\ref{lay9_DMHE3}) are the subsystem model constraints, while Equations (\ref{lay9_DMHE4}) and (\ref{lay9_DMHE5}) consider the constraints on system states, measurement noise, and system disturbances. Compared with other distributed MHE, the difference of the algorithm proposed in this paper is that Equations (\ref{lay9_DMHE6}) and (\ref{lay9_DMHE7}) take into account the constraints of the variable selection results for each subsystem. Here, let $U(t)$ denote the set containing the indices of the unselected elements of $\bfx_{\theta}(t)$ based on the variable selection presented in Subsection \ref{orthogonalization}. If $x_{\theta,3}$, $x_{\theta,7}$ and $x_{\theta,8}$ are not selected, then $U(t)=\{3,7,8\}$. This means that the elements in subsystem disturbance vector corresponding to the unselected variable are zero, that is, $\bfw^i_{\theta,u}=0$ ($u\in U(t)$). In this case, the unselected variables $\tilde{\bfx}^i_{\theta,u}$ will only evolve in an open-loop manner according to the subsystem model, and the initial condition $\tilde{\bfx}^i_{\theta,u}(t-N)$ is designated as the value obtained at the previous instant.

For the proposed distributed framework, each MHE estimator is required to exchange information with each other. At each sampling time, each local MHE is performed to provide the state estimates of the subsystem based on the information gathered from its associated subsystem and its interactive subsystems. The detail steps of the DMHE are shown as follows:
\begin{enumerate}
\item At time $t=0$, initialize the MHE of each subsystem by using a initial guess $\hat{\bfx}^i_{\theta}(0)$, and set the sampling data length $L$.

\item At time $t>0$, each MHE receives the output measurements of corresponding subsystem, and receives the state estimates of other subsystems at previous time instant.

\item Calculate the state estimates of each MHE according the received information.

\item If $t<L$, increase $t$ by 1 and go to Step 2; otherwise, obtain the augmented state estimates $\hat{\bfx}^i_{\theta}(t)$ of each system.
\end{enumerate}

\section{Application to a chemical process}
\label{Examples}

\subsection{Process description}
\begin{figure}[!hbt]
 \centering
 \includegraphics[width=0.8\hsize]{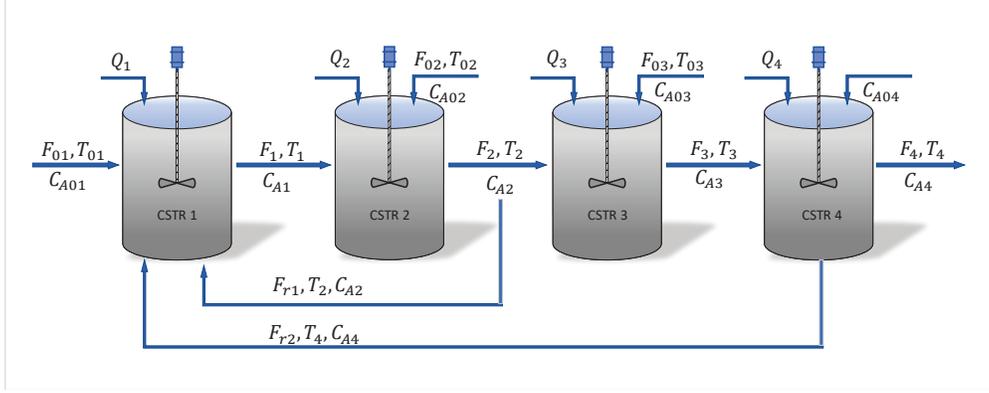}
 \caption{A schematic of the four-CSTR process}
\label{lsy9_fig_4CSTR}
\end{figure}

In this section, we apply the proposed method to a chemical process consisting of four connected continuous-stirred tank reactors (CSTRs) with different volumes \cite{Rashedi2018}. The four vessels are interconnected with each other via material and energy flows. A schematic of the process is shown in Figure \ref{lsy9_fig_4CSTR}.

The pure reactant $A$ is fed into the $i$th vessel for reactions at temperature $T_{0i}$, flow rate $F_{0i}$, and molar concentration $C_{0i}$, $i=1,2,3,4$. During the reaction process, the outlet flow of the first three vessels is supplied to the adjacent downstream tank, respectively. For the second vessel, besides constituting the feed stream of the third vessel, a part of its outlet stream is recycled back to the first vessel at flow rate $F_{r1}$, temperature $T_2$ and molar concentration $C_2$. A part of effluent from the fourth vessel is recycled back to the first vessel at flow rate $F_{r2}$, temperature $T_4$ and molar concentration $C_4$, and the other part is the discharge stream for further processing. The heat is removed from/provided to the corresponding vessel by attaching a jacket to each CSTR. Based on mass and energy balances, a model that comprises eight differential equations describes the process dynamics:
\begin{align}
 \frac{dC_{A1}}{dt}=&\frac{F_{01}}{V_1}(C_{A01}-C_{A1})+\frac{F_{r1}}{V_1}(C_{A2}-C_{A1})\non\\
       &+\frac{F_{r2}}{V_1}(C_{A4}-C_{A1})-\sum^3_{i=1}k_i e^{\frac{-E_i}{RT_1}}C_{A1},\\
 \frac{dT_1}{dt}=&\frac{F_{01}}{V_1}(T_{01}-T_1)+\frac{F_{r1}}{V_1}(T_2-T_1)+\frac{F_{r2}}{V_1}(T_4-T_1)\non\\
       &-\sum^3_{i=1}\frac{\Delta H_i}{\rho c_p}k_i e^{\frac{-E_i}{RT_1}}C_{A1}+\frac{Q_1}{\rho c_p V_1},\\
 \frac{dC_{A2}}{dt}=&\frac{F_1}{V_2}(C_{A1}-C_{A2})+\frac{F_{02}}{V_2}(C_{A02}-C_{A2})\non\\
       &-\sum^3_{i=1}k_i e^{\frac{-E_i}{RT_2}}C_{A2},\\
 \frac{dT_2}{dt}=&\frac{F_1}{V_2}(T_1-T_2)+\frac{F_{02}}{V_2}(T_{02}-T_2)\non\\
       &-\sum^3_{i=1}\frac{\Delta H_i}{\rho c_p}k_i e^{\frac{-E_i}{R T_2}}C_{A2}+\frac{Q_2}{\rho c_p V_2},\\
 \frac{dC_{A3}}{dt}=&\frac{F_2-F_{r1}}{V_3}(C_{A2}-C_{A3})+\frac{F_{03}}{V_3}(C_{A03}-C_{A3})\non\\
       &-\sum^3_{i=1}k_i e^{\frac{-E_i}{RT_3}}C_{A3},\\
 \frac{dT_3}{dt}=&\frac{F_2-F_{r1}}{V_3}(T_2-T_3)+\frac{F_{03}}{V_3}(T_{03}-T_3)\non\\
       &-\sum^3_{i=1}\frac{\Delta H_i}{\rho c_p}k_i e^{\frac{-E_i}{R T_3}}C_{A3}+\frac{Q_3}{\rho c_p V_3},\\
 \frac{dC_{A4}}{dt}=&\frac{F_3}{V_4}(C_{A3}-C_{A4})+\frac{F_{04}}{V_4}(C_{A04}-C_{A4})\non\\
       &-\sum^3_{i=1}k_i e^{\frac{-E_i}{RT_4}}C_{A4},\\
 \frac{dT_4}{dt}=&\frac{F_3}{V_4}(T_3-T_4)+\frac{F_{04}}{V_4}(T_{04}-T_4)\non\\
       &-\sum^3_{i=1}\frac{\Delta H_i}{\rho c_p}k_i e^{\frac{-E_i}{R T_4}}C_{A4}+\frac{Q_4}{\rho c_p V_4},
\end{align}
where $C_{Ai}$ and $T_i$ are the molar concentration of $A$ and temperature in the $i$-th vessel, $i=1,2,3,4$. The definitions of the other variables are presented in Table~\ref{lsy9_tab1}. Table~\ref{lsy9_tab2} shows the values of model parameters. For this process, the heating inputs $Q_i$ to the four vessels are chosen to be the input variable, $C_{Ai}$ and $T_i$ are the system states, where $T_i$ is measured online. Each measurement is associated with a subsystem, and a local estimator is developed based on the measurement. The continuous model is discretized using the fourth-order Runge-Kutta method with a sample time $\Delta t=\frac{1}{120}$h.

\begin{table}[!hbt] \small
 \centering
 \caption{Other process variables of four-CSTR model}
 \label{lsy9_tab1}
\renewcommand{\arraystretch}{1.2}
 \tabcolsep 1pt
 \begin{tabular}{ll}\hline
 Variables & Defination  \\\hline
  $V_i$    & Liquid holdup volume in the $i$-th vessel   \\
  $F_i$    & Flow rate of the effluent of the $i$-th vessel   \\
  $E_1$, $E_2$, $E_3$ & Activation energies\\
  $\Delta H_1$, $\Delta H_2$, $\Delta H_3$ & Enthalpies of reactions\\
  $k_1$, $k_2$, $k_3$ & Pre-exponential constants for reactions\\\hline
\end{tabular}\end{table}

\begin{table}[!hbt] \small
 \centering
 \caption{Parameter values of the four-CSTR process}
 \label{lsy9_tab2}
\renewcommand{\arraystretch}{1.2}
 \tabcolsep 6pt
 \begin{tabular}{ll}\hline
  \multicolumn{2}{l}{Known parameters}   \\\hline
  $T_{01}=300$ K    &  $\Delta H_1=-5.0\times 10^4$ kJ/kmol  \\
  $T_{03}=300$ K    &  $\Delta H_2=-5.2\times 10^4$ kJ/kmol  \\
  $T_{02}=300$ K    &  $\Delta H_3=-5.0\times 10^4$ kJ/kmol  \\
  $T_{04}=300$ K    &  $k_1=3.0\times10^6$ $\rm h^{-1}$  \\
  $c_p=0.231$ kJ/(kg$\cdot$K)  & $k_2=3.0\times10^5$ $\rm h^{-1}$ \\
  $\rho=1000$ kg/m$^3$         & $k_3=3.0\times10^5$ $\rm h^{-1}$ \\\hline
  \multicolumn{2}{l}{Unknown parameters to be estimated}   \\\hline
  $F_{01}=5$ $\rm m^3$/h    & $C_{01}=4.0$ kmol/$\rm m^3$  \\
  $F_{02}=10$ $\rm m^3$/h   & $C_{02}=2.0$ kmol/$\rm m^3$  \\
  $F_{03}=8$ $\rm m^3$/h    & $C_{03}=3.0$ kmol/$\rm m^3$  \\
  $F_{04}=12$ $\rm m^3$/h   & $C_{04}=3.5$ kmol/$\rm m^3$  \\
  $V_1=1$ $\rm m^3$         & $F_1=35$ $\rm m^3$/h         \\
  $V_2=3$ $\rm m^3$         & $F_2=45$ $\rm m^3$/h         \\
  $V_3=4$ $\rm m^3$         & $F_3=33$ $\rm m^3$/h         \\
  $V_4=6$ $\rm m^3$         & $F_{r1}=20$ $\rm m^3$/h      \\
  $E_1=5.0\times10^4$ kJ/kmol  & $F_{r2}=10$ $\rm m^3$/h   \\
  $E_2=7.5\times10^4$ kJ/kmol  & $R=8.314$ kJ/(kmol$\cdot$K) \\
  $E_3=7.53\times10^4$ kJ/kmol & \\\hline
 \end{tabular}\end{table}

\subsection{Augmented system construction and variable selection}
According to the dynamic model of the four-CSTR process, there are 8 state variables and 21 unknown parameters, we need to estimate these variables by using the proposed variable selection, community structure detection, and distributed moving horizon estimation methods. The first step is to construct the augmented system. The state vector $\bfx=[C_{A1}, T_1, C_{A2}, T_2, C_{A3}, T_3, C_{A4}, T_4]^{\T}$ and parameter vector $\bftheta=[F_{0i}, V_i, C_{0i}, E_1, E_2, E_3, F_1, F_2, F_3, F_{r1}, F_{r2}, R]^{\T}$ ($i=1,2,3,4$) constructs the augmented state as follows:
\[
 \bfx_{\theta}:=[\bfx^{\T}, \bftheta^{\T}]^{\T}.
\]
According to the parameter values in Table~\ref{lsy9_tab2}, the four-CSTR process has a steady state
\begin{align}
 \bfx_s=&[2.78\rm kmol/m^3, 363K, 2.58kmol/m^3, 356K, 2.6\rm kmol/m^3, 355K, 2.6kmol/m^3, 392K]^{\T}.\non
\end{align}
Next, we normalize the augmented model around the steady state $\bfx_s$ and the nominal parameter values in Table~\ref{lsy9_tab2} to avoid the potential impact of the parameter value in rank calculation.

In the simulations, the constant heat inputs to four vessels are selected as: $Q_{1}=1.0\times10^4$kJ/h, $Q_{2}=2.0\times10^4$kJ/h, $Q_{3}=2.5\times10^4$kJ/h, and $Q_{4}=1.0\times10^4$kJ/h. In the variable selection, a predetermined cut-off value is required to terminate the variable selection process. Here, we propose to use the following cut-off value \cite{Liu2021_IECR}:
\begin{equation}
\label{Exam_1}
 \alpha=3\sqrt{\bar{\bfomega}^2+\bar{\bfv}^2},
\end{equation}
where $\bar{\bfomega}=10^{-3}$ and $\bar{\bfv}=10^{-3}$ denote the standard deviation of normalized process noise and measurement noise, respectively. In this way, the influence of noises on each element in the normalized model is basically at the same level. It is challenging to know that the model contains exactly how much noise information in the sensitivity matrix. The cut-off value $\alpha$ in (\ref{Exam_1}) is an empirical expression used to approximate the noise level in this work. Other methods of determining the cut-off value of variable selection are also applicable.

In the variable selection, we consider the eight original states are important and should be estimated at each sampling time. Therefore, the variable selection is only performed among the parameters. The information represented by eight states needs to be removed from the sensitivity matrix $\bfS_{\theta}(k)$. Table~\ref{lsy9_tab3} shows the number of sampling times of each state and parameter (The total sampling times are 500). It can be seen that nine parameters are selected based on the sensitivity information and will be estimated together with the states.

\begin{table*}[!hbt] \small
 \centering
 \caption{Number of sampling times for each variable }
 \label{lsy9_tab3}
\renewcommand{\arraystretch}{1.2}
 \begin{tabular}{lccccccccccccccc}\hline
         & $C_{A1}$ & $T_1$ & $C_{A2}$ & $T_2$ & $C_{A3}$ & $T_3$ & $C_{A4}$ & $T_4$ & $F_1$ & $F_2$ & $F_3$ & $V_1$ & $V_2$ & $V_3$  & $V_4$\\\hline
  Count & 499&500&499&500&499&500&499&500&0&0&0&498&498&498&498\\\hline
  & $F_{r1}$ & $F_{r2}$ & $E_1$ & $E_2$ & $E_3$ & $R$ & $F_{01}$ & $F_{02}$ & $F_{03}$ & $F_{04}$ & $C_{01}$ & $C_{02}$ & $C_{03}$ & $C_{04}$&\\\hline
  Count &0&497&0&0&0&0&497&497&497&496&0&0&0&0&\\\hline
\end{tabular}\end{table*}

\subsection{Subsystem decomposition}
In the community structure detection, the initial values of $\Omega$ and $m$ are set to zero. Apply the proposed method to decompose these variables obtained by the sensitivity analysis. The best structure is case 2 in Table~\ref{lsy9_tab4}, obtained by maximum $\Omega$ ($\Omega$=0.4497), which divides the system into three sub-systems. To verify the effectiveness of the proposed method, three other cases are considered. In case 1, we use the centralized MHE to estimate the most appropriate 17 variables in Table~\ref{lsy9_tab3} selected by the variable selection algorithm based on the sensitivity analysis. In case 3, instead of using the community structure detection to decompose the four-CSTR system after the variable selection, it is divided into four subsystems ($\Omega$=0.3892) based on intuitive conjecture as shown in Table~\ref{lsy9_tab4}. To verify the effectiveness of sensitivity analysis, all the 29 variables in the augmented state are estimated in case 4, without the sensitivity analysis and community structure detection. That means the insensitive variables are assigned to the three subsystems in case 2.

\begin{table}[!hbt] \small
 \centering
 \caption{Subsystem description for cases 2, 3, and 4}
 \label{lsy9_tab4}
\renewcommand{\arraystretch}{1.2}
 \tabcolsep 6pt
 \begin{tabular}{lccc}\hline
  Case 2 ($\Omega$=0.4497)  & States      & Inputs  \\\hline
  sub 1   &$C_1$, $\bf{T_1}$, $C_2$, $\bf{T_2}$, $F_{01}$, $F_{02}$, $V_1$, $V_2$, $F_{r2}$  & $Q_1$, $Q_2$ \\
  sub 2   &$C_3$, $\bf{T_3}$, $F_{03}$, $V_3$    &  $Q_3$ \\
  sub 3   &$C_4$, $\bf{T_4}$, $F_{04}$, $V_4$    &  $Q_4$\\\hline
  Case 3 ($\Omega$=0.3892) & States      & Inputs  \\\hline
  sub 1   &$C_1$, $\bf{T_1}$, $F_{01}$, $V_1$, $F_{r2}$  & $Q_1$  \\
  sub 2   &$C_2$, $\bf{T_2}$, $F_{02}$, $V_2$    &  $Q_2$ \\
  sub 3   &$C_3$, $\bf{T_3}$, $F_{03}$, $V_3$    &  $Q_3$ \\
  sub 4   &$C_4$, $\bf{T_4}$, $F_{04}$, $V_4$    &  $Q_4$\\\hline
  Case 4  & States      & Inputs  \\\hline
  sub 1   &$C_1$, $\bf{T_1}$, $C_2$, $\bf{T_2}$, $F_{01}$, $F_{02}$, $V_1$, $V_2$,  & $Q_1$, $Q_2$ \\
          &$C_{01}$, $C_{02}$, $F_1$, $F_{r1}$, $F_{r2}$, $R$, $E_1$       &     \\
  sub 2   &$C_3$, $\bf{T_3}$, $F_{03}$, $V_3$, $C_{03}$, $F_2$, $E_2$    &  $Q_3$ \\
  sub 3   &$C_4$, $\bf{T_4}$, $F_{04}$, $V_4$, $C_{04}$, $F_3$, $E_3$    &  $Q_4$\\\hline
\end{tabular}\end{table}

\subsection{Estimation results}
\begin{figure}[!hbt]
 \centering
 \includegraphics[width=0.8\hsize]{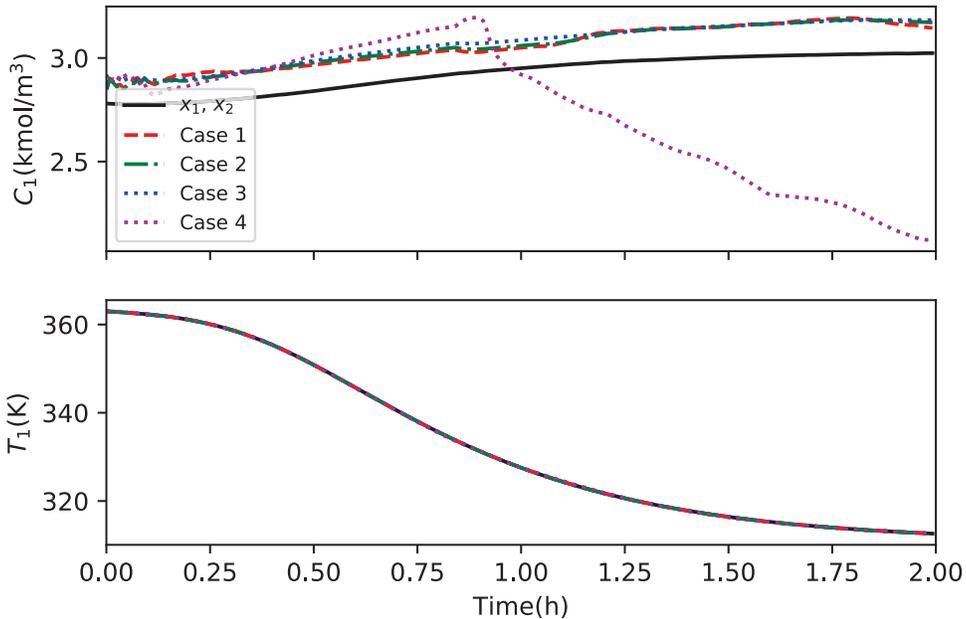}
 \caption{Trajectories of the actual states $C_1$, $T_1$ (solid lines), and state estimates in cases 1, 2, 3, and 4 (dashed lines, dash-dot lines, blue dotted lines, pink dotted lines)}
\label{lsy9_fig2}
\end{figure}

\begin{figure}[!hbt]
 \centering
 \includegraphics[width=0.8\hsize]{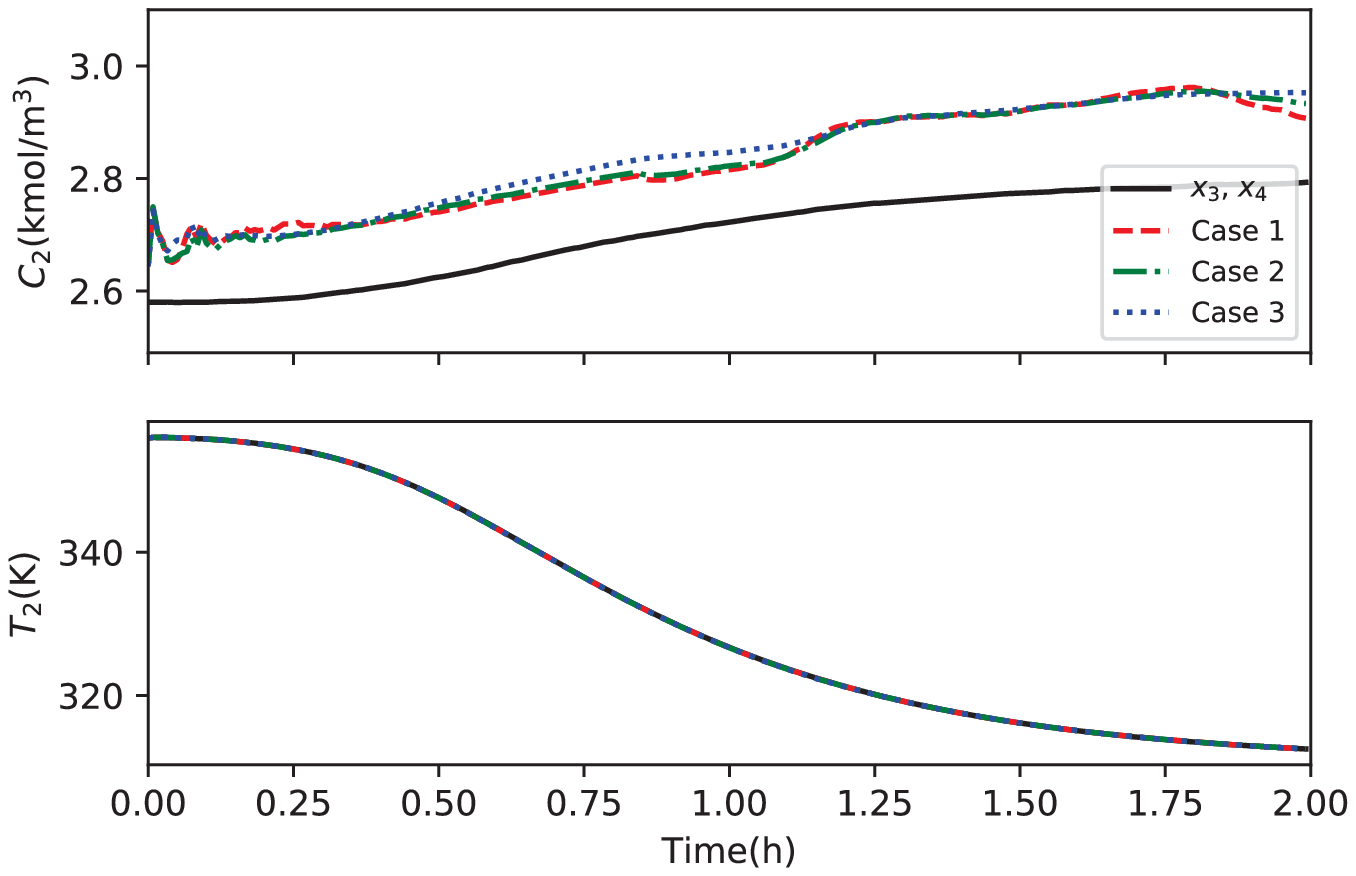}
 \caption{Trajectories of the actual states $C_2$, $T_2$, and state estimates in cases 1, 2, and 3}
\label{lsy9_fig3}
\end{figure}

\begin{figure}[!hbt]
 \centering
 \includegraphics[width=.8\hsize]{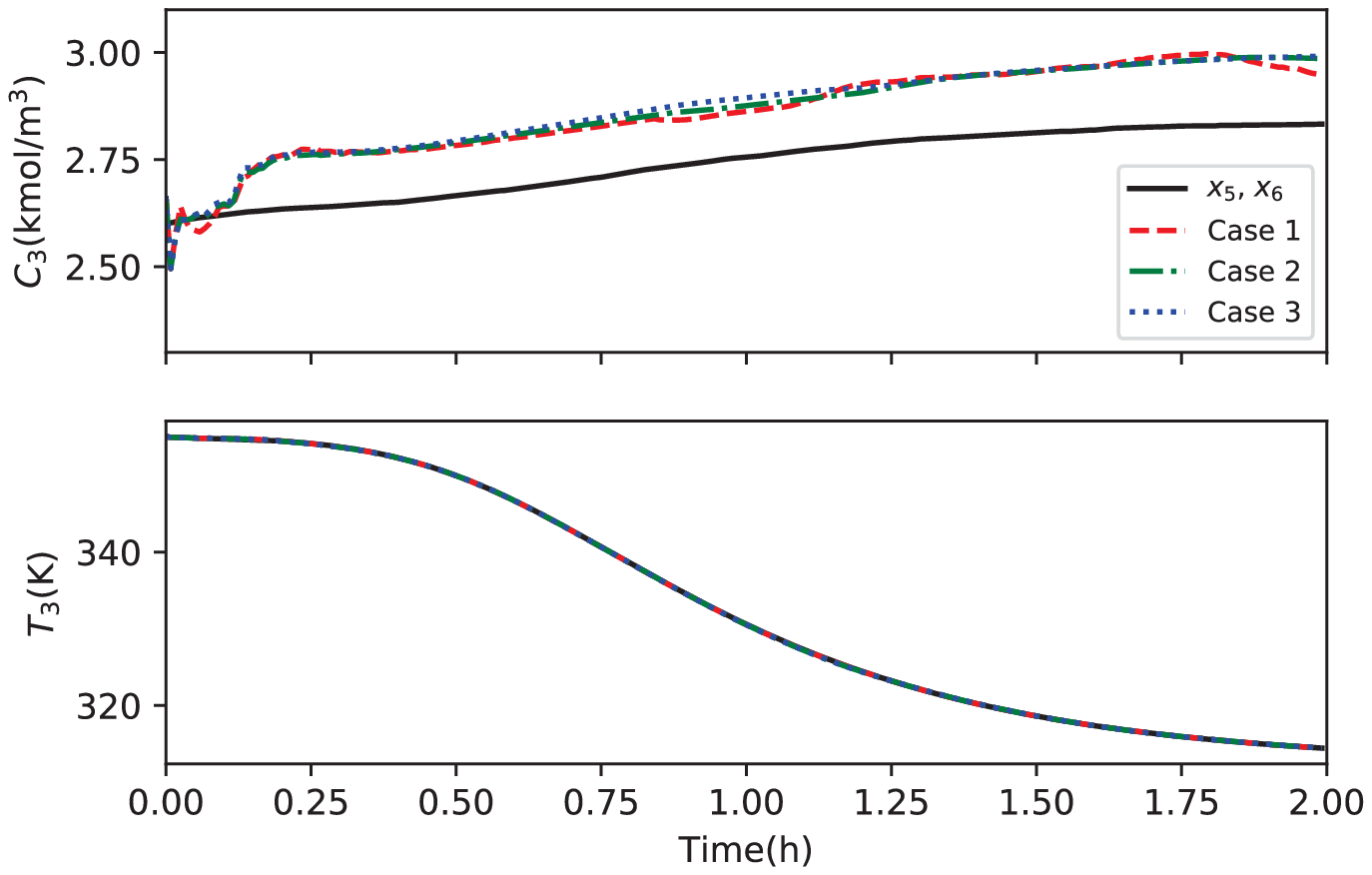}
 \caption{Trajectories of the actual states $C_3$, $T_3$, and state estimates in cases 1, 2, and 3}
\label{lsy9_fig4}
\end{figure}

\begin{figure}[!hbt]
 \centering
 \includegraphics[width=.8\hsize]{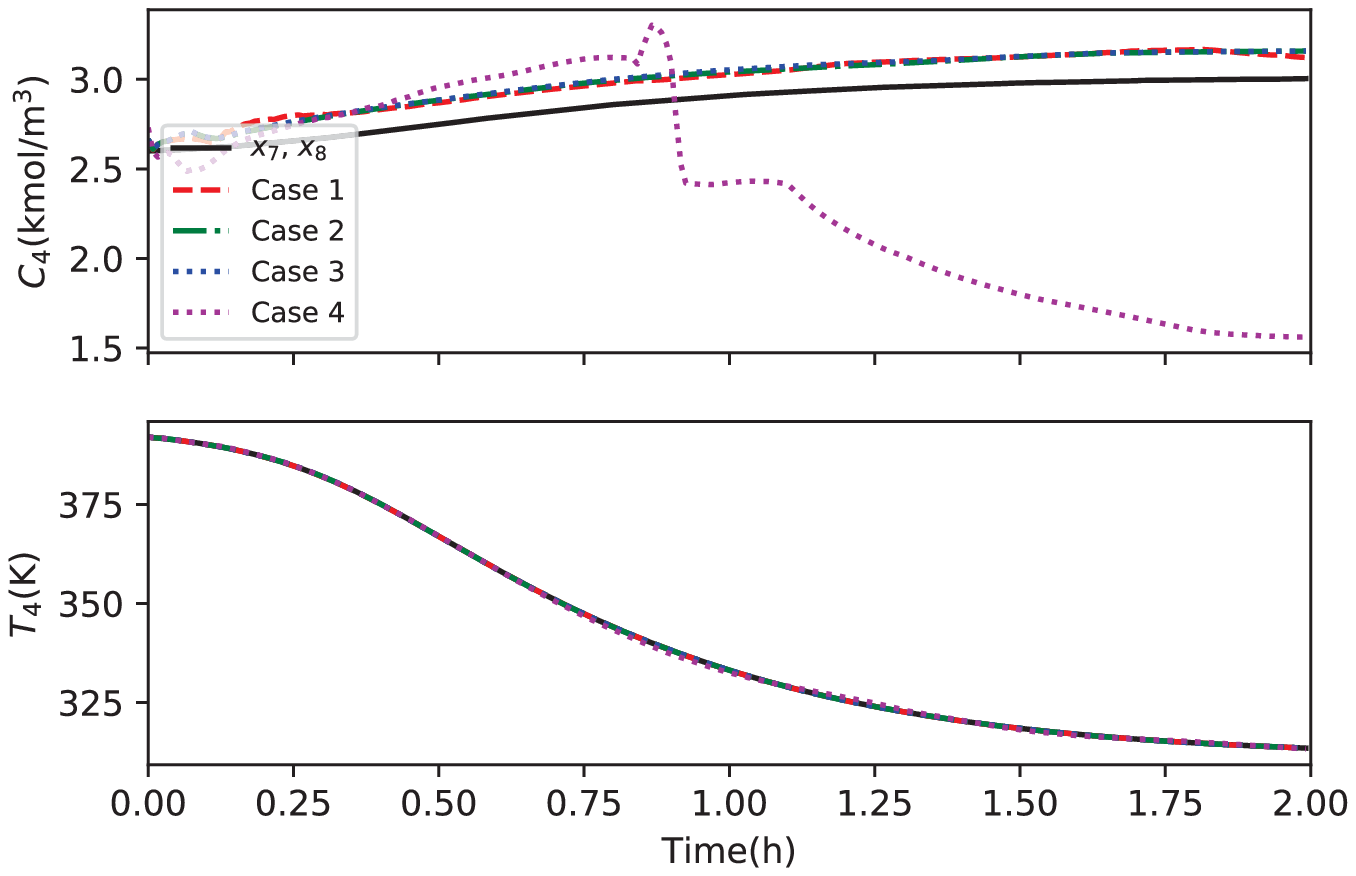}
 \caption{Trajectories of the actual states $C_4$, $T_4$, and state estimates in cases 1, 2, 3, and 4}
\label{lsy9_fig5}
\end{figure}

\begin{figure}[!hbt]
 \centering
 \includegraphics[width=0.8\hsize]{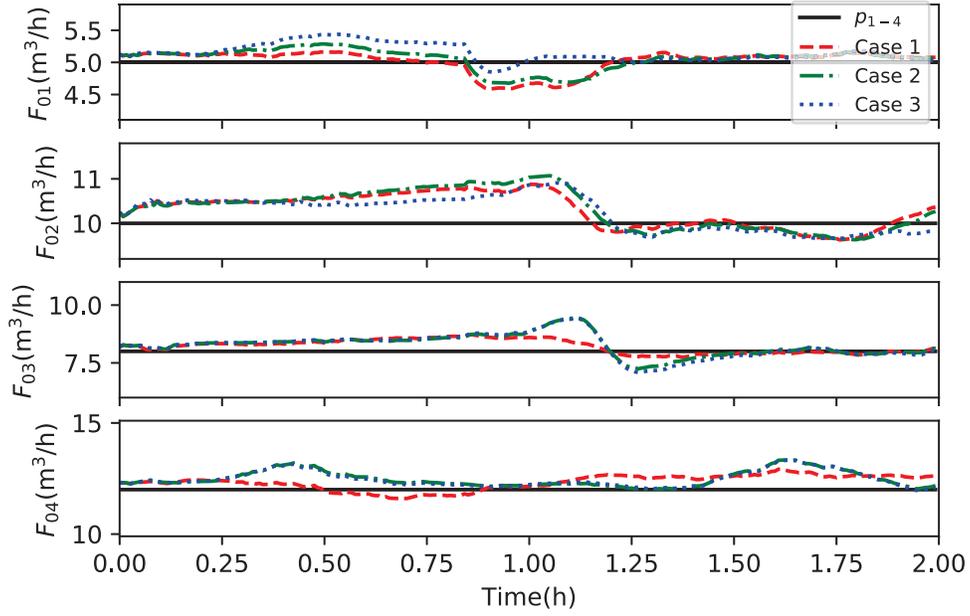}
 \caption{Trajectories of the actual parameters $F_{01}$, $F_{02}$, $F_{03}$, $F_{04}$, and parameter estimates in cases 1, 2, and 3}
\label{lsy9_fig6}
\end{figure}

\begin{figure}[!hbt]
 \centering
 \includegraphics[width=0.8\hsize]{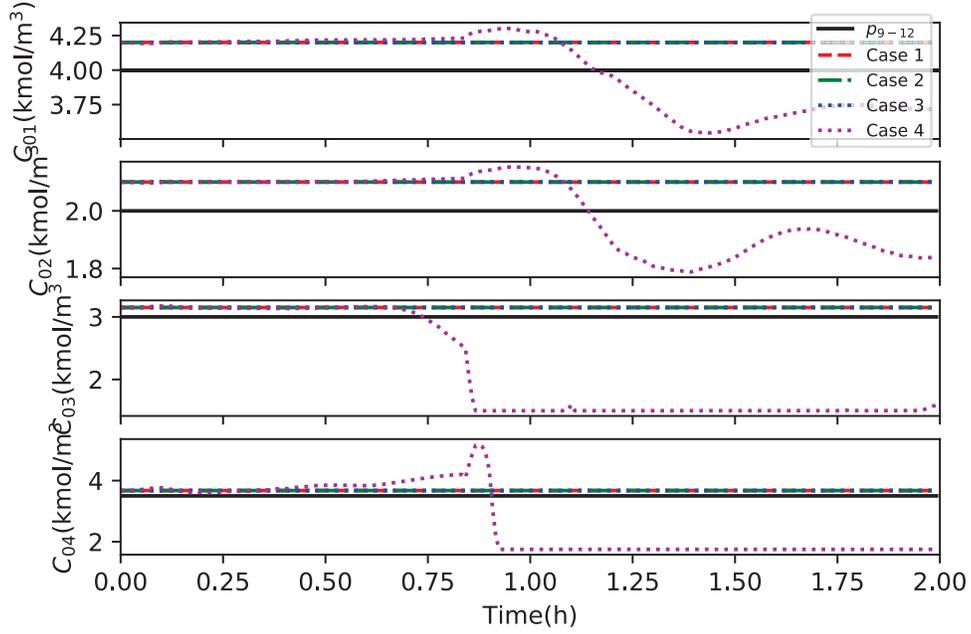}
 \caption{Trajectories of the actual parameters $C_{01}$, $C_{02}$, $C_{03}$, $C_{04}$, and parameter estimates in cases 1, 2, 3, and 4}
\label{lsy9_fig7}
\end{figure}

\begin{figure}[!hbt]
 \centering
 \includegraphics[width=0.8\hsize]{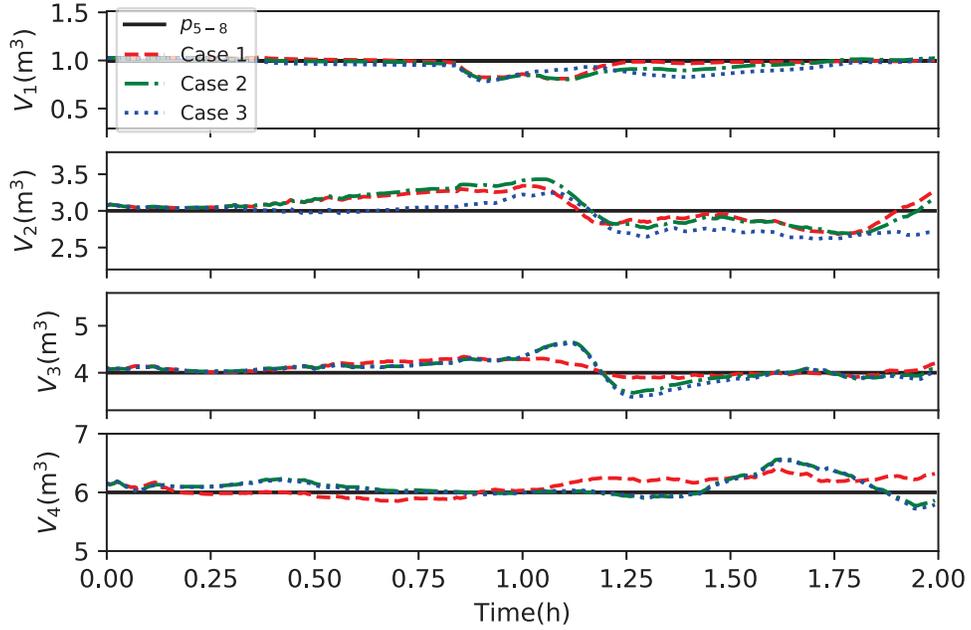}
 \caption{Trajectories of the actual parameters $V_1$, $V_2$, $V_3$, $V_4$, and parameter estimates in cases 1, 2, and 3}
\label{lsy9_fig8}
\end{figure}

\begin{figure}[!hbt]
 \centering
 \includegraphics[width=0.8\hsize]{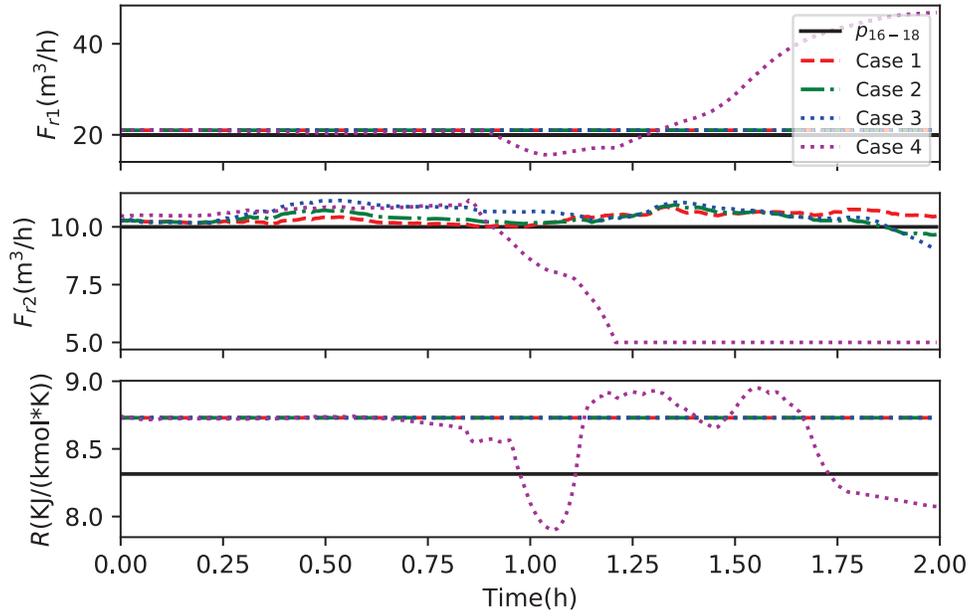}
 \caption{Trajectories of the actual parameters $F_{r1}$, $F_{r2}$, $R$, and parameter estimates in cases 1, 2, 3, and 4}
\label{lsy9_fig9}
\end{figure}

\begin{figure}[!hbt]
 \centering
 \includegraphics[width=0.8\hsize]{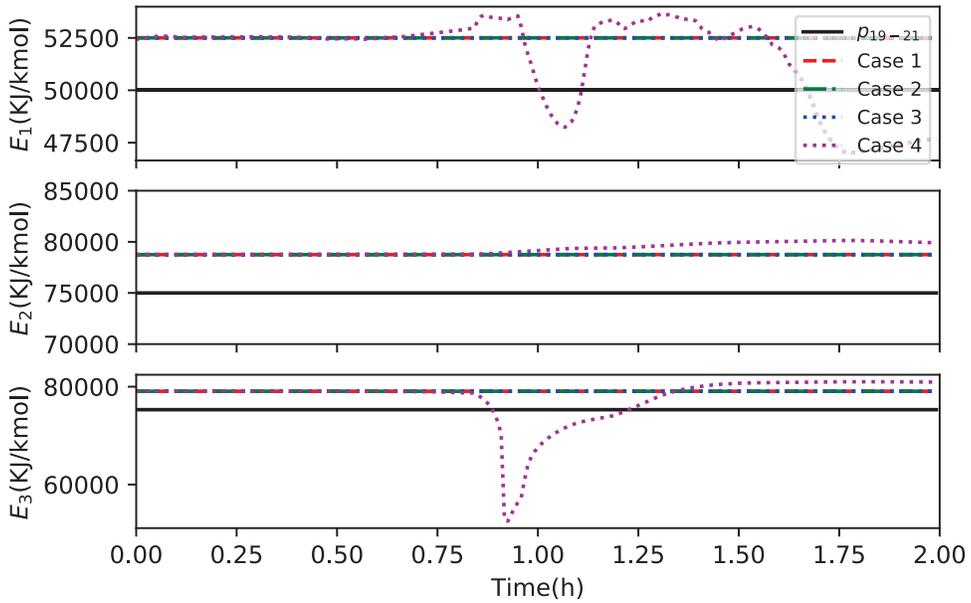}
 \caption{Trajectories of the actual parameters $E_1$, $E_2$, $E_3$, and parameter estimates in cases 1, 2, 3, and 4}
\label{lsy9_fig10}
\end{figure}

\begin{figure}[!hbt]
 \centering
 \includegraphics[width=0.8\hsize]{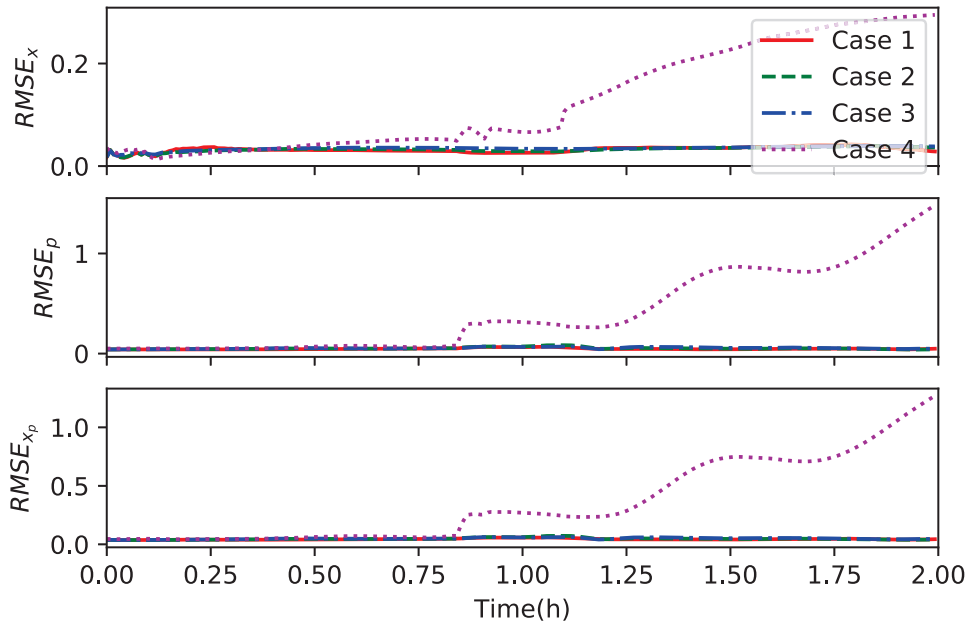}
 \caption{Evolution of the RMSE of the eight original state vector, the parameter vector, and the entire augmented state vector }
\label{lsy9_fig11}
\end{figure}

In both centralized and distributed MHE simulation design, a 5\% mismatch in the initial state of each of the eight original states is considered. It is also assumed that the parameters are not known exactly and there is a 5\% mismatch in each of the parameters. The covariance parameters of each MHE are tuned as $\bfQ_1={\rm diag}\{0.05^2, 0.05^2, 0.05^2, 0.05^2\}$, $\bfQ_2=\bfQ_3={\rm diag}\{0.05^2, 0.05^2\}$, and $\bfR_1={\rm diag}\{0.05^2, 0.05^2\}$, $R_2=R_3=0.05^2$, and the matrix $\bfP_1={\rm diag}\{0.1^2\bfI_4, 0.07^2\bfI_5\}$, $\bfP_2=\bfP_3={\rm diag}\{0.1^2\bfI_2, 0.07^2\bfI_2\}$. The four cases use the same input and output data. The simulation results of four cases are shown in Figures~\ref{lsy9_fig2}--\ref{lsy9_fig11}. From Figures~\ref{lsy9_fig2}--\ref{lsy9_fig10}, it can be seen that the state estimation performance of case 4 is much poorer compared with cases 1, 2, and 3. The main reason is that the rank of the sensitivity matrix of augmented system is 17, which is smaller than 29. This means that ignoring the observability when estimating the augmented state vector $\bfx_{\theta}$ will cause bad estimation results. In cases 1, 2, and 3, the observability is considered, and a subset of variables selected based on the sensitivity matrix is estimated. The estimation performance of CMHE with sensitivity analysis in case 1 is best. The estimation performance of the proposed SA and CSD based DMHE in case 2 is similar to that of case 1. It means that the proposed distributed strategy is feasible. For case 3, the estimation performance is inferior to that of case 2, which shows that the method of the community structure detection is effective.

\begin{remark}
In order to show the differences among cases 1, 2, and 3 more clearly, we choose to remove the estimation results of the case 4 in Figures 3, 4, 6, and 8, because the estimation error in case 4 is so large that the curves of the first three cases almost overlap.
\end{remark}

To assess the estimation performance, the root-mean-square error (RMSE) at a time instant and the average RMSE as the performance indexes are applied, which is defined as
\begin{align}
 {\rm RMSE}_{x_{\theta}}(k)&:=\sqrt{\frac{\sum^{n_{x_{\theta}}}_{j=1}[(\bfx_{\theta,j}(k)-\hat{\bfx}_{\theta,j}(k))/\bfx_{\theta,j}(k)]^2}{n_{x_{\theta}}}},\non\\
 {\rm RMSE}_{x_{\theta}}&:=\frac{\sum^{N-1}_{k=0}{\rm RMSE}_{x_{\theta}}(k)}{N},\non
\end{align}
where $N$ indicates the total simulation time/steps, $\bfx_{\theta,j}$ is the true states, and $\hat{\bfx}_{\theta,j}$ is the state estimate of $j$-th element in the augmented state. ${\rm RMSE}_{x_{\theta}}(k)$ denotes the evolution of the RMSE value over time, $k=0,1,\cdots,N-1$, and ${\rm RMSE}_{x_{\theta}}$ denotes the average value. Figure~\ref{lsy9_fig11} shows ${\rm RMSE}_x$ for the original states, ${\rm RMSE}_{\theta}$ for all parameters, and ${\rm RMSE}_{x_{\theta}}$ for entire augmented state over time $k$. The average RMSE of four cases is shown in Table~\ref{lsy9_tab5}. It can be found that the RMSE value of case 4 is higher than the others. The results of case 1 are the best, followed by case 2, and then case 3.

\begin{table}[!hbt] \small
 \centering
 \caption{Average RMSEs of four cases}
 \label{lsy9_tab5}
\renewcommand{\arraystretch}{1.2}
 \tabcolsep 8pt
 \begin{tabular}{lcccc}\hline
                               & Case 1 & Case 2 & Case 3 & Case 4 \\\hline
  ${\rm RMSE}_x(\%)$           & 3.23   & 3.26   & 3.45   & 12.65  \\
  ${\rm RMSE}_{\theta}(\%)$    & 4.80   & 5.19   & 5.42   & 41.06  \\
  ${\rm RMSE}_{x_{\theta}}(\%)$& 4.44   & 4.76   & 4.96   & 35.64  \\\hline
\end{tabular}\end{table}



\section{Conclusion}
\label{Section 6}

In this work, a systematic procedure is proposed for distributed simultaneous state and parameter estimation for nonlinear systems when the corresponding augmented models are note fully observable. The proposed method resorts to sensitivity analysis based on the orthogonalization to test the observability for nonlinear systems and to select variables for simultaneous estimation. In the proposed procedure, the community structure detection based on modularity maximization lays a foundation of distributed simultaneous estimation. A fast unfolding algorithm that approximately maximizes the modularity is used to find candidate subsystem configurations. The results of a chemical process example demonstrate the effectiveness and applicability of the proposed algorithm.

\section{Acknowledgement}

The first author, S.Y. Liu, is a visiting Ph.D. student in the Department of Chemical and Materials Engineering at the University of Alberta from March 2021 to February 2023. She acknowledges the financial support from the China Scholarship Council (CSC) during this period.


\end{document}